\documentstyle[sprocl]{article}
\def\be{\begin{equation}}
\def\ee{\end{equation}}
\def\bea{\begin{eqnarray}}
\def\eea{\end{eqnarray}}

\begin{document}

\title{DARK MATTER PARTICLES}

\author{V.Berezinsky}

\address{INFN, Laboratori Nazionali del Gran Sasso, 67010 Assergi (AQ), 
Italy\\ and Institute for Nuclear Research, Moscow}
\maketitle

\abstracts{
The baryonic and cold dark matter are reviewed in the context of 
cosmological models. The theoretical search for the particle candidates 
is limited by supersymmetric extension of the Standard Model. Generically
in such models there are just two candidates  associated with each other: 
generalized neutralino, which components are usual neutralino and axino, 
and axion which is a partner of axino in supermultiplet. The status of these 
particles as DM candidates is described.}

\section{\bf Introduction}

Presence of dark matter (DM) in the Universe is reliably established.
DM in the form of compact microlensing objects (machos) 
are directly observed in the halo of our Galaxy by MACHO, EROS and OGLE 
collaborations. 
Rotation curves in our Galaxy and in many other galaxies provide evidence for 
large halos 
filled by nonluminous matter. The virial (gravitational) mass of clusters 
of galaxies is about ten times larger than their luminous masses.
IRAS and POTENT 
demonstrate the presence of DM on the largest scale in the Universe. 

The matter density in the Universe $\rho$ is usually parametrized in 
terms of $\Omega=\rho/\rho_c$, where 
$\rho_c\approx 1.88\cdot10^{-29}h^2~g/cm^3$ is the critical density and 
$h$ is the dimensionless Hubble constant defined as 
$h=H_0/(100 km.s^{-1}.Mpc^{-1})$. Different measurements suggest generally
$0.4\leq h \leq 1$. The recent measurements of extragalactic 
Cepheids in Virgo and Coma clusters narrowed this interval to 
$0.6 \leq h \leq 0.9$. However, one should be cautious about the accuracy of 
this interval due to uncertainties involved in these difficult measuremets.

Dark Matter can be subdivided in baryonic DM, hot DM (HDM) and 
cold DM (CDM).

The density of baryonic matter found from nucleosynthesis
is given \cite{CST95} as $0.009 \leq \Omega_b h^2 \leq 0.02.$ The baryonic 
cosmological density provided by the mass of intracluster gas is very close 
to this value, $\Omega_b^{clust}h^{3/2} \approx 0.05$ \cite{WhF95} 
and references therein.  

The structure formation in Universe put strong  restrictions to  the properties 
of DM in Universe. Universe with HDM plus baryonic DM  has 
a wrong prediction for the spectrum of fluctuations as compared with 
measurements of COBE, IRAS and CfA. CDM plus baryonic matter can explain the 
spectrum of fluctuations if total density $\Omega_0 \approx 0.3$.
   
There is one more form of energy density in the Universe, namely the 
vacuum energy described by the cosmological constant $\Lambda$. The 
corresponding energy density is given by
$\Omega_{\Lambda}= \Lambda/(3H_0^2)$. Quasar lensing restricts the vacuum   
energy density: in terms of $\Omega_{\Lambda}$ it is less than 0.7 \cite{Koch93}
.

Contribution of galactic halos to the total density is estimated as 
$\Omega \sim 0.03 - 0.1$ and clusters give $\Omega \approx 0.3$.
Inspired mostly by theoretical motivation (horizon problem, flatness problem 
and the beauty of the inflationary scenarios) $\Omega_0=1$ is usually assumed. 
This value is supported by IRAS data and POTENT analysis. No observational data
 significantly contradict this value.

There are several cosmological models based on the four types of DM
described above (baryonic DM, HDM, CDM and vacuum energy). These models 
predict different spectra of fluctuations to be compared with data of  
COBE, IRAS, CfA etc. They also produce  different
effects for cluster-cluster correlations, velocity dispersion etc. The  
simplest and most attractive 
model for a correct description of all these phenomena is the so-called
mixed model or  cold-hot dark matter model (CHDM). This model is
characterized by following parameters:
\begin{eqnarray}
\Omega_{\Lambda}=0, \Omega_0=\Omega_b+\Omega_{CDM}+\Omega_{HDM}=1,\nonumber\\
H_0\approx 50~km s^{-1}Mpc^{-1} (h \approx 0.5), \nonumber\\
\Omega_{CDM}:\Omega_{HDM}:\Omega_b \approx 0.75:0.20:0.05,
\label{eq:chdm}
\end{eqnarray}
Thus in the CHDM model the central value for the CDM density  
is given by $\Omega_{CDM}h^2 = 0.19$, with uncertaities within 0.1.

The best candidate for the HDM particle is $\tau$-neutrino. In the CHDM 
model with $\Omega_{\nu}=0.2$ mass of $\tau$ neutrino is
$m_{\nu_{\tau}} \approx 4.7~eV$. This component will not be discussed further.

The most plausible candidate for the CDM particle is probably the neutralino 
($\chi$):
it is massive, stable (when the neutralino is the lightest supersymmeric 
particle and if R-parity is conserved) and the $\chi\chi$-annihilation 
cross-section results in $\Omega_{\chi} h^2 \sim 0.2$ in  large areas of 
the neutralino parameter space.

In the light of recent measurements of the Hubble constant the CHDM model 
faces the {\em age problem}.
The lower limit on the age of Universe $t_0 > 13$~Gyr (age of globular
clusters) imposes the upper limit on the Hubble constant in the CHDM model 
$H_0 < 50~km s^{-1}Mpc^{-1}$. This value is in slight contradiction with the
recent observations of extragalactic Cepheids,
which can be summarized as $H_0 > 60~km s^{-1} Mpc^{-1}$.
However, it is too early to speak about a serious conflict taking into account
the many uncertainties and the physical possibilities (e.g. the Universe can be 
locally overdense - see the discussion in ref.\cite{Pri95}).

The age problem, if to take it seriously, can be solved with help of 
another successful cosmological model $\Lambda$CDM. This model assumes that 
$\Omega_0=1$ is provided by the vacuum energy described by cosmological 
constant $\Lambda$ and CDM.
Using the limit on cosmological constant $\Omega_{\Lambda}<0.7$ and the age of
Universe one obtains  $\Omega_{CDM} \geq 0.3$ and 
$h<0.7$. Thus this model also predicts
$\Omega_{CDM} h^2 \approx 0.15$ with uncertainties 0.1. Finally, we shall 
mention that the CDM with $\Omega_0=\Omega_{CDM}=0.3$ and $h=0.8$, which 
fits the observational data, also gives $\Omega h^2 \approx 0.2$. Therefore 
$\Omega h^2 \approx 0.2 \pm 0.1$ can be considered as the value common for most 
models \cite{NT}.

We shall analyze here the candidates for CDM which naturally arise from 
elementary particle physics. The best known solution for strong CP 
violation implies {\em axion}, which can serve as CDM particle. The 
supersymmetrization of the theory, which includes axion, results in 
{\em generalized neutralino} -- 
a linear combination of five neutral spin 1/2 particles (wino, bino,
two higgsinos and {\em axino}, the fermionic partner of axion). This 
generalized neutralino is most natural candidate for CDM particle. 
And finally some attention will be given to the baryonic DM in connection 
with observations of machos.

\section{Machos and Baryonic Dark Matter}
The total number of microlensing events observed in the halo during last two 
years 
reached 10. Eight of them are observed by MACHO collaboration \cite{MACHO1}
and two -- by 
EROS. The duration of lensing effect 
is determined by the lens mass. The distribution of observed durations 
yields the macho mass as \cite{MACHO1} $M=0.46^{+0.30}_{-0.17} M_{\odot}$.
However, this value is model dependent.
The most likely halo fraction of machos is \cite{MACHO1} 
$f=0.50^{+0.30}_{-0.20}$. The important result is observation
\cite{MACHO2} of 45 microlensing events in Galactic bulge. For a given 
rotation curve the heavy bulge implies the lighter halo and thus the 
{\em fraction} of machos increases for a given number of observed events.

The machos with these masses should be interpreted as white dwarfs. 
However,
to escape from the Hubble Deep Field Search these objects must be very 
faint, two magnitudes fainter than the disc white dwarf sequence 
\cite{Hubble}.  

The DM in the halo of our Galaxy is found and most probably it is baryonic.
Could be all DM in the Universe only baryonic? This question is often asked 
nowdays (e.g. see \cite{Paol}).

Let us discuss shortly the problems arising in the baryonic-dominated\\ 
Universe.

Nucleosynthesis requires $\Omega_B^{nucl} \leq 0.02 h^{-2}$. On the other 
hand the clusters provide $\Omega_{DM}^{cl} \geq 0.2$. Therefore,
the baryonic density is small ($\Omega_B^{nucl} < \Omega_{DM}^{cl}$),
unless $h < 0.3$, which contradicts recent observations. If one arbitrary 
neglects this contradiction, the baryonic dominated Universe with
$\Omega_B=\Omega_{DM}^{cl} \approx 0.2$ can be considered. Apart from 
IRAS data and POTENT analysis which give $\Omega \approx 1$, such model 
faces serious cosmological problems, including the horizon and flatness 
problem and observed spectrum of fluctuations, which is impossible to 
explain without CDM and HDM. Probably, these problems could be solved in 
some artificial models with $\Lambda$ term and vacuum defects (e.g. 
strings), but at present the corresponding calculations do not exist.

As was mentioned above, the baryonic nature of machos have (or can have)
the problems. An interesting idea about the nature of machos was recently 
put forward in ref.\cite{Gur}. These objects could be the neutralino stars,
the formations produced by neutralinos and baryons around singularities
\cite{GZ} 
in the distribution of neutralino gas. The neutralino stars are produced 
naturally and they do not meet any problems connected with the Hubble 
telescope observations. Unfortunatelly, as was demonstrated in 
ref.\cite{ns} these objects produce too high gamma-ray flux due to 
annihilation of neutralinos.  

\section{Axion}
The axion is generically a light pseudoscalar particle which gives natural and  
beautiful 
solution to the CP violation in the strong interaction \cite{PQ77}
(for a review and references see\cite{KT90}). Spontaneous breaking of the 
PQ-symmetry due to VEV of the scalar field  $<\phi>= f_{PQ}$ results in the 
production of massless Goldstone boson. Though $f_{PQ}$ is a free 
parameter, in practical applications it is assumed to be large,
$f_{PQ} \sim 10^{10} - 10^{12}~GeV$ and therefore the PQ-phase transition 
occurs in very early Universe. At low temperature 
$T \sim \Lambda_{QCD} \sim 0.1~GeV$ the chiral anomaly 
of QCD induces the mass of the Goldstone boson
$m_a \sim \Lambda_{QCD}^2/f_{PQ}$ . This massive Goldstone particle is the 
{\em axion}. The interaction of axion is basically determined by the Yukawa 
interactions of field(s) $\phi$ with fermions. Triangular anomaly, which 
provides the axion mass, results in the coupling of the axion with two photons.
Thus, the basic  for cosmology and astrophysics axion interactions are those 
with nucleons, electrons and photons. 

Numerically, axion mass is given by
\be
m_a=1.9\cdot 10^{-3}(N/3)(10^{10}~GeV/f_{PQ})~eV,
\label{eq:am}
\ee
where $N$ is a color anomaly (number of quark doublets).

All coupling constants of the axion are inversely proportional to $f_{PQ}$
and thus are determined by the axion mass. Therefore, the upper limits on 
emission 
of axions by stars result in upper limits for the axion mass. In Table 1 we 
cite  the upper limits 
on axion mass from ref.\cite{KT90}, compared with revised limits, given 
recently by Raffelt \cite{Raf95}.
\begin{table}[hbt]
\caption{Astrophysical upper limits on axion mass}
\center{\begin{tabular}{|c|c|c|} \hline
                    &1990 \cite{KT90}  & 1996 \cite{Raf95}\\
\hline
sun                  &$1~eV$                & $1~eV$\\ \hline

red giants           &$1\cdot 10^{-2}~eV$   & \begin{tabular}{l}
                                                very \\
                                               uncertain
                                            \end{tabular} \\ \hline
\begin{tabular}{l}
hor.--branch \\
stars 
\end{tabular}  & \begin{tabular}{l}
                  not \\
                  considered
                  \end{tabular} 
& $0.4~eV$\\ \hline
SN 1987A             &$1\cdot 10^{-3}~eV$   & $1\cdot 10^{-2}~eV$\\
\hline
\end{tabular}}
\end{table}

As one can see from the Table the strong upper limit, given in 1990 from
red giants, is replaced by the weaker limit due to the horizontal-branch 
stars. The upper limit from SN 1987A was reconsidered taking into account 
the nucleon spin fluctuation in $N+N \to N+N+a$ axion emission.

There are three known mechanisms of {\em cosmological production} of 
axions.  They are (i)thermal production, (i) misalignment production and 
(iii) radiation from axionic strings.

The relic density of thermally produced axions is about the same as 
for light neutrinos and thus for 
the mass of axion $m_a \sim 10^{-2}~eV$ this component is not important 
as DM.

The {\em misalignment production} is clearly explained in ref.\cite{KT90}.

At very low temperature $T\ll\Lambda_{QCD}$ the massive axion provides 
the minimum of the potential at value $\theta=0$,which corresponds to 
conservation of CP. At very high temperatures $T\gg \Lambda_{QCD}$ the axion 
is massless and the potential does not depend on $\theta$.  At these 
temperatures 
there is no reason for 
$\theta$ to be zero: its values are different in various casually 
disconnected regions of the Universe. When $T\to\Lambda_{QCD}$ the system 
tends to go to potential minimum (at $\theta=0$) and as a result oscillates 
around this position. The energy of these coherent oscillations is
the axion energy density in the Universe. From cosmological point of view 
axions in this regime are equivalent to CDM. The energy density of this 
component is approximately \cite{KT90}$^{\!,\,}$\cite{Bat94}

\be
\Omega_ah^2 \approx 2\cdot (m_a/10^{-5}~eV)^{-1.18}.
\ee
Uncertainties of the calculations can be estimated as $10^{\pm 0.5}$.

Axions can be also produced by radiation of {\em axionic strings}
\cite{KT90,BaSh94}. Axionic string is a one-dimension vacuum defect 
$<\phi_{PQ}>=0$, i.e. a line of old vacuum embedded into the new one. 
The string network includes the long strings and closed loops which 
radiate axions due to oscillation. There were many uncertainties in the 
axion radiation by axionic strings (see ref.\cite{KT90} for a review). 
Recently more detailed and accurate calculations were performed by 
Battye and Schellard \cite{BaSh94}. They obtained for the density of axions
\be
\Omega_ah^2 \approx A(m_a/10^{-5}~eV)^{-1.18}
\ee
with $A$ limited between 2.7 and 15.2 and 
with uncertainties of the order $10^{\pm 0.6}$. The overproduction condition 
$\Omega_ah^2 >1$ imposes lower limit on axion mass 
$m_a> 2.3\cdot10^{-5}~eV$.
\vspace*{90mm}
\begin{figure}[htb]
\caption{Axion window 1996.
The curves "therm." and "misalign." describe the thermal and misalignement
production of axions, respectively. The dash-dotted curve corresponds to the
calculations by Davis \protect \cite{Dav86} for string production. 
The recent refined calculations \protect \cite{BaSh94} are shown by two dashed 
lines for two extreme cases, respectively. The other explanations are given 
in the text.}
\end{figure}
Fig.1 shows the density of axions $\Omega_a h^2$ as a function of the 
axion mass $m_a$. The upper limits  on axion mass from Table I are shown 
above the upper absciss (limits of 1990) and below lower absciss (limits 
of 1996). The overproduction region $\Omega_ah^2>1$ and the regions excluded 
by astrophysical observations \cite{Raf95} are shown as the dotted areas.

The axion window of 1996 (shown as undotted region) 
became wider and moved to the right as compared with window 1990.
The  horizontal strip shows $\Omega_{CDM}=0.2 \pm 0.1$ as it was 
discussed in Introduction. One can see from Fig.1 that string and 
misaligment mechanisms provide the axion density as required by 
cosmological CDM model, if axion mass is limited between 
$7\cdot10^{-5}~eV$ and $7\cdot 10^{-4}~eV$. However, in the light of 
 uncertainties, mostly in the calculations of axion production,
one can expect that this "best calculated" window is between
$3\cdot 10^{-5}$ and $10^{-3}$~eV. This region is partly 
overlapped with a possible direct search for the axion in nearest-future 
experiments (see Fig.1 and refs.\cite{axdet}).

\section{Generalized neutralino}

We shall consider now the supersymmetric theories where strong CP-violation
is suppressed by PQ-symmetry (note, however, that in supersymmetric theories  
the new mechanisms for suppression of CP violation might appear, see 
e.g. \cite{Moh}).
In supersymmetric theory the PQ symmetry breaking results in the 
production of the 
Goldstone chiral supermultiplet which contains two scalar fields and their 
fermionic partner -- axino ($\tilde{a}$). The scalar fields enter the 
supermultiplet 
in the combination $(f_{PQ}+s)\exp(a/f_{PQ})$, where $s$ is a scalar field,
saxino, which describes the oscillations of the initial field $\phi$ around 
its VEV value $<\phi>=f_{PQ}$, and $a$ is the axion field. This phase
transition in the Universe occurs at temperature $T \sim f_{PQ}$. 
As we saw in the previous section the axion is massless at this temperature 
and since supersymmetry is not broken yet, the axino and saxino are massless,
too. The axion acquires the mass in the usual way due to chiral anomaly at 
$T\sim \Lambda_{QCD}$, while saxino and axino obtain the masses due to 
global supersymmetry breaking. 

The saxino is not of great interest for 
cosmology: it is heavy and it decays fast (mostly into two gluons).

In the Minimal Supersymmetric Standard Model (MSSM) with broken PQ symmetry we 
have 
five spin 1/2 neutral particles: wino $\tilde{W}_3$, bino $\tilde{B}$, two 
Higgsinos ($\tilde{H_1}$ and $\tilde{H_2}$) and axino $\tilde{a}$. 
Generically the Lightest Supersymmetric Particle (LSP) is a linear 
superposition of these 5 fields:
\be
\tilde{\chi}=C_1\tilde{W}_3+C_2\tilde{B}+C_3\tilde{H}_1+C_4\tilde{H}_2+
C_5\tilde{a}
\label{eq:gn} 
\ee
Further on we shall consider two extreme cases: generalized neutralino 
$\tilde{\chi}$ is strongly dominated by the axino state and it is dominated
by the first four terms in eq.(\ref{eq:gn}). In the former case LSP is 
almost pure axino, in the latter-- the usual neutralino. 

\section{Axino}

How heavy the axino can be?
The mass of axino is very model dependent.
In the phenomenological approach, using the global supersymmetry breaking 
parameter $M_{SUSY}$ one typically obtains (e.g. \cite{TW82,Nie86}) 
\be
m_{\tilde{a}} \sim M_{SUSY}^2/f_{PQ}
\label{eq:ma}
\ee
For example, if global SUSY breaking occurs due to VEV of auxiliary field of
the goldstino supermultiplet $<F>=F_g$, then the axino mass appears due to 
interaction term $(g/f_{PQ})\tilde{a}\tilde{a}F$ (F has a dimension $M^2$),
and using $<F>=F_g=M_{SUSY}^2$ one arrives at the value (\ref{eq:ma}).

The situation is different in supergravity. In ref.\cite{ChL95} the general
analysis of the axino mass is given in the framework of local supersymmetry.
It was found that generically the mass of axino in these theories is 
$m_{\tilde{a}} \sim m_{3/2} \sim 100~GeV$. Even in case when  axino 
mass is small at tree level, the radiative corrections raise this mass 
to the value $\sim m_{3/2}$. This result holds for the most general form of 
superpotential.The global SUSY result,
$m_{\tilde{a}} \sim m_{3/2}^2/f_{PQ}$, can be reproduced in the local SUSY
only if one of the superpotential coupling constants is very small,
$\lambda <10^{-4}$, which implies fine-tuning. Thus, the axino is too heavy 
 to be a CDM particle.

The only exceptional case was found by 
Goto and Yamaguchi \cite{GY92}. They demonstrated that in case of no-scale  
superpotential the axino mass vanishes and the radiative corrections in 
some specific models can result in the axino mass $10 - 100~keV$, cosmologically
interesting. This beautiful case gives essentially the main foundation for 
axino as CDM particle.

The cosmological production of axinos can occur through thermal production
\cite{RTW91}
or due to  decays of the 
neutralinos \cite{BGM89,RTW91}. The axion chiral supermultiplet 
contains two particles which can be CDM particles, namely axion and 
axino. In this section we are interested in the case when axino gives the 
dominant contribution. In particular this can take place in the range  
$2\cdot 10^9~GeV <f_{PQ}< 2.7\cdot 10^{10}~GeV$ where axions are cosmologically 
unimportant.

Since axino interacts with  matter very weakly, the decoupling temperature 
for the thermal production is very high \cite{RTW91}:
\be 
T_d \approx 10^9~GeV(f_{PQ}/10^{11}~GeV).
\ee
Therefore, axinos are produced thermally at the reheating phase after
inflation. The relic concentration of axinos can be easily evaluated 
for the reheating temperature $T_R$ as
\be
\Omega_{\tilde{a}}h^2 \approx 0.6 \frac{m_{\tilde{a}}}{100~keV}
(\frac{3\cdot 10^{10}~GeV}{f_{PQ}})^2\frac{T_R}{10^9~GeV}
\label{eq:omega}
\ee
Reheating temperature $T_R \leq 10^9~GeV$ gives no problem with the gravitino 
production. The relic density (\ref{eq:omega}) provides 
$\Omega_{CDM}h^2 \sim 0.2$ for a reasonable set of parameters $m_{\tilde{a}},
f_{PQ}$ and $T_R$.

If the axino is LSP and the neutralino is the second lightest supersymmetric 
particle, the axinos can also be produced by neutralino 
decays \cite{BGM89,RTW91,BGM94}. According to estimates 
of ref.\cite{BGM94} the axinos are produced due to 
$\chi \to \tilde{a}+\gamma$ decays at the epoch with red-shift 
$z_{dec}\sim 10^8$. Axinos are produced in these decays as ultrarelativistic 
particles and the free-streeming prevents the growth  of fluctuations on the 
horizon scale and less. At red-shift $z_{nr} \sim 10^4$ axinos become
non-relativistic due to adiabatic expansion (red shift). From this moment on 
the axinos behave as the usual CDM and the fluctuations on the scales 
$\lambda \geq (1+z_{nr})ct_{nr}$ (which correspond to a mass larger than 
 $ 10^{15} M_{\odot}$) grow as in the case of standard CDM.
For smaller scales the fluctuations, as was explained above, grow less
than in CDM model. Therefore, as was observed in ref.\cite{BGM94}, the 
axinos produced by neutralino decay behave like HDM. It means that
axinos
can provide generically both components, CDM and HDM, needed for description of 
observed spectrum of fluctuations.

Unfortunatelly stable axino is unobservable. In case of very weak R-parity 
violation, decay of axinos can produce a diffuse X-ray radiation, with 
practically no signature of the axino.

\section{Neutralino}
The generalized neutralino can be dominated by the first four 
terms in eq.(\ref{eq:gn}) 
\be
\chi = C_1\tilde{W}_3+C_2\tilde{B}+C_3\tilde{H}_1+C_4\tilde{H}_2
\label{eq:chi}
\ee
i.e. by usual neutralino.

The neutralino is a Majorana particle. With a unitary relation between the
coefficients $C_i$ the parameter space of neutralino states is described by
three independent parameters, e.g. mass of wino $M_2$, mixing parameter of two 
Higgsinos $\mu$, and the ratio of two vacuum expectation values 
$\tan\beta=v_2/v_1$.

In literature one can find two extreme approaches 
describing the neutralino as a DM particle.

(i){\em Phenomenological approach}.
The allowed neutralino parameter space is restricted by the LEP and CDF data.
In particular these data put a lower limit to the neutralino mass,
$m_{\chi}> 20$~GeV. In this approach only the usual GUT relation between  
 gaugino masses, $M_1:M_2:M_3=\alpha_1:\alpha_2:\alpha_3$, 
is used  as an additional assumption, where $\alpha_i$
are the gauge coupling constants. All other SUSY masses which are needed for the 
calculations are treated
as free parameters, limited from below by accelerator data. 

One can find the relevant 
calculations within this approach in refs.\cite{Bott94,Bott95}
and in the review\cite{JKG} (see  also the references therein). There are 
large areas
in neutralino parameter space where the neutralino relic density satisfies  
$\Omega_{CDM}h^2 \approx 0.2 \pm 0.1$. This is especially true for heavy 
neutralinos with
$m_{\chi}>100 - 1000$~GeV, ref.\cite{Tur}. In these areas there are  good
prospects for {\em indirect} detection of neutralinos, due to high energy 
neutrino radiation from Earth and Sun (see \cite{Ka91,Bott95a} and
references therein) as well as due to production of antiprotons and positrons 
in our  Galaxy. The {\em direct} detection of neutralinos is possible too, 
though in more restricted parameter space areas of light neutralinos 
(see review \cite{JKG}).

(ii) {\em Strongly constrained models}.
This approach is based on  the remarkable observation that in the minimal 
SUSY SU(5) model with fixed particle content, the three running coupling
constants meet at one point corresponding to the GUT mass $M_{GUT}$. 
Because of the fixed particle content of the model, its predictions are 
rigid and they strongly restrict the neutralino parameter space.
This is especially true for the limits due to proton decay 
$p \to K^{+}\nu$. As a result very little space is left for neutralino as DM
particle. Normally neutralinos overclose the Universe ($\Omega_{\chi}>1$). 
The relic density decreases to the allowed values in very restricted  areas 
where  $\chi\chi$-annihilation is accidentally 
large (e.g.due to the $Z^0$ exchange term - see 
ref.\cite{Nan92}. Thus, this approach looks rather pessimistic for 
neutralino as DM particle.

In several recent works \cite{Nan93}$^{\!-\,}$\cite{RS95} less restricted SUSY
models were considered with more optimistic conclusions about detection 
prospects.

(iii){\em Relaxed restrictions}.
In some recent works the restrictions described in (ii) are relaxed. 
In particular, in \cite{BG95} the large number of models with relaxed 
conditions were analysed. It was found that for many models neutralino can
be discovered in the direct and indirect detection experiments. In refs.
\cite{Ber95}$^{\!,\,}$\cite{NT} the SUSY models only with basic 
restrictions were considered.
  
\section{\bf SUSY models with  basic restrictions}
Following refs. \cite{Ber95,NT} 
we shall consider here the  restrictions to 
neutralino as DM particle, imposed by {\em basic} properties of SUSY theory.
These restrictions are as  follows:\\
\noindent
(i)Radiative Electroweak Symmetry Breaking (EWSB), which is 
considered as 
fundamental element of the analysis,  (ii) No fine-tuning stronger than $1\%$, 
which is natural but very powerful requirement; it results in the upper limit 
to neutralino mass $m_{\chi}<200~GeV$,
(iii) Restrictions from Renormalization Group Equations (RGE) and from 
particle phenomenology (accelerator limits on the calculated masses and 
the condition that neutralino is LSP), (iv) Limits from $b \to s\gamma$ decay
taken with the uncertainties in the calculations of the decay rate 
and (v) $0.01 < \Omega_{\chi}h^2<1$ as the allowed relic density for 
neutralinos.

At the same time some restrictions are lifted as being too model-dependent:
(i) No restrictions are imposed due to $p \to K\nu$ decay, (ii) Unification 
of coupling 
constants at the GUT point is allowed to be not  exact (it is 
assumed that new very heavy particles can restore the unification),
(iii) unification in the soft breaking terms is relaxed.
Following ref.\cite{Pok95} it is assumed that masses of Higgses at the GUT 
scale can deviate from the universal  value $m_0$ as
\be
m_{H_i}^2(GUT)=m_0^2(1+\delta_i)\qquad(i=1,2).
\ee

 This non-universality affects rather strongly the properties of neutralino 
as DM particle: the allowed parameter space regions become larger and 
neutralino is allowed to be Higgsino-dominated, which is favorable for 
detection.

Some results obtained in ref.\cite{Ber95}$^{\!,\,}$\cite{NT} are illustrated by 
Figs.~2 - 3.

In Fig.2 the regions excluded by the LEP and 
CDF data are shown by dots and labelled as LEP. The regions labelled 
"fine tuning" have an accidental compensation stronger than $1\%$
and thus are excluded.  No-fine-tuning
region inside the broken-line box corresponds to a neutralino mass 
$m_{\chi}\leq 200$~GeV. The region "EWSB+particle phenom." is excluded
by the EWSB condition
combined with particle phenomenology (neutralino as LSP, limits on the 
masses of SUSY particles etc). In the region marked by rarefied dotted lines 
neutralinos overclose the Universe ($\Omega_{\chi}h^2>1$). The solid line 
corresponds to  $m_0=0$. The regions allowed for neutralino 
as CDM particle ($0.01<\Omega_{\chi}h^2<1$) are shown by small boxes. As 
one can see in most regions the neutralinos are overproduced. The allowed regions 
correspond to large $\chi\chi$ annihilation cross-section (e.g. due to 
$Z^0$-pole).

Fig.~2a and Fig.~2b differ only by universality: in Fig.~2a 
$\delta_1=\delta_2=0$ (mass--unification), while in Fig.~2b $\delta_1=-0.2$
and $\delta_2=0.4$. The allowed region in Fig.~2b
becomes much larger and is shifted into the Higgsino dominated region.
\newpage
\vspace*{80mm}
\begin{figure*}[h]
\caption{ The neutralino parameter space for {\bf(a)}\hspace{1mm}
mass--unification case
$\delta_1=\delta_2=0$  and \hspace{1mm}{\bf(b)}\hspace{1mm} for non-- 
universal case $\delta_1=-0.2$, $\delta_2=0.4$. Both cases are given for 
$\tan\beta=8$.}
\end{figure*}

Let us discuss now the predictions of this model for direct and indirect 
detection of neutralinos. Direct detection is based on observarions of
recoil nuclei from neutralino-nucleus scattering. As indirect detection 
we shall consider here the registration of high energy neutrinos from 
neutralino-neutralino annihilation in the center of Earth and Sun.

In Fig.~3a the scatter plot for the rate of direct detection with the 
$Ge$ detector \cite{Beck94} is given for the non-universal 
case ($\delta_1=0, \delta_2=-0.2$) and 
$\tan\beta=53$. We notice that, for some configurations, the experimental
sensitivity \cite{Beck94} is already at the level of the predicted rate.

In Fig.~3b we show predictions for the updoing flux of muons produced 
by neutrinos from neutralino-neutralino annihilation in the core of Earth.
The muon flux from the direction of the Sun is shown in Fig.~3c.  The 
horizontal solid curves in both cases  present the observational  upper 
limits \cite{Bol95}.
One can see that in examples given above the muon fluxes can be reliably 
detected by future gigantic neutrino telescopes.

\vspace*{80mm}
\begin{figure}[htb]
\caption{Direct (a) and indirect (b and c) detection of neutralinos for 
$\tan\beta=53$ and different neutralino masses. Fig.3a presents counting rate
in units $events/(kg\cdot day)$.  Fig.3b gives the 
underground muon flux from the Earth-core direction for 
$\delta_1=\delta_2=0$ (solid line), $\delta_1=0, \delta_2=-0.3$ 
(dashed line), and  $\delta_1=0.7, \delta_2=0.4$ (dotted line).
Fig.3c gives the underground muon flux from the direction of the Sun 
for the case $\delta_1=0$ and $\delta_2=-0.3$. ; the fluxes are given in 
units $cm^{-2}s^{-1}$.
}
\end{figure}

\section{\bf Conclusions}

The baryonic DM is discovered in the halo of our Galaxy. Machos have a mass
between $0.1 - 1.2 M_{\odot}$ and comprise between $10 - 100 \%$ of the 
total mass of galactic halo. However, the bulk of DM observed in the 
Universe can hardly be dominated by baryons.

The most successful cosmological models require CDM with 
density $\Omega h^2 \approx 0.2\pm 0.1$. The minimal supersymmetric 
extension of SM, with strong CP violation suppressed by PQ-symmetry, predicts
the generalized neutralino as a superposition of usual neutralino and 
axino. The Goldstone chiral supermultiplet contains axion and axino. 
Therefore, in this model  there are three natural candidates for
CDM particle: axion, axino and neutralino (or a linear superposition of the 
latter two).

The new {\em axion} window corresponds to axion masses between $3\cdot 10^{-5}$
 and 
$3\cdot 10^{-3}~eV$, i.e. it only partly overlaps with the range of search in 
microwave cavity experiments.
   
{\em Axino} can provide both CDM and HDM. The direct observation of this 
particle seems to be impossible.

{\em Neutralino} remains most attractive CDM candidate. In the models with 
radiative EW symmetry breaking the properties of neutralino are 
restricted, but there are many configurations where the neutralino can 
provide the required $\Omega_{CDM}h^2$ and can be found by direct and indirect 
methods.

{\bf Acknowledgements} \hfill \break
The main results presented here on the neutralino are based on a work
carried out with Sandro Bottino, John Ellis, Nicolao Fornengo, Guilio 
Mignola and Stefano Scopel. I  wish to 
express many thanks to them for collaboration and for discussion of other 
topics included in this talk.

\end{document}